\documentclass{aa}

\usepackage{natbib}
\usepackage{graphicx}
\usepackage{txfonts}

\begin{document}
\title{Structure of the SMC}
\subtitle{Stellar component distribution from 2MASS data}
\author{
I. Gonidakis \inst{1}
\and E. Livanou \inst{1}
\and E. Kontizas \inst{2}
\and U. Klein \inst{3}
\and M. Kontizas \inst{1}
\and M. Belcheva \inst{1}
\and P. Tsalmantza \inst{4}
\and A. Karampelas \inst{1}
}

\offprints{E. Livanou\\
\email{elivanou@phys.uoa.gr}}

\institute{Department of Astrophysics Astronomy \& Mechanics,
 Faculty of Physics, University of Athens, GR-15783 Athens, Greece
\and Institute for Astronomy and Astrophysics, National
 Observatory of Athens, P.O. Box 20048, GR-11810 Athens, Greece
\and Radioastronomisches Institut der Universitat Bonn,
 Auf dem Hogel 71, D-53121 Bonn, Germany
\and Max-Planck-Institut fur Astronomie, Konigstuhl 17,
 69117 Heidelberg, Germany
}

\date{Received date / accepted}

\abstract {} {The spatial distribution of the SMC stellar component
is investigated from 2MASS data. The morphology of the different age
populations is presented. The center of the distribution is
calculated and compared with previous estimations. The rotation of
the stellar content and possible consequence of dark matter presence
are discussed.} {The different stellar populations are identified
through a CMD diagram of the 2MASS data. Isopleth contour maps are
produced in every case, to reveal the spatial distribution. The
derived density profiles are discussed.} {The older stellar
population follows an exponential profile at projected diameters of
about 5 kpc ($\sim$5$^{o}$) for the major axis and  $\sim$ 4 kpc for
the minor axis, centred at RA: 0$^{h}$ 51$^{min}$, Dec: -73$^{o}$
7$\arcmin$ (J2000.0). The centre coordinates are found the same for
all the different age population maps and are in good accordance
with the kinematical centre of the SMC. However they are found
considerably different from the coordinates of the centre of the gas
distribution. The fact that the older population found on an
exponential disk, gives evidence that the stellar content is
rotating, with a possible consequence of dark matter presence. The
strong interactions between the MCs and the MilkyWay might explain
the difference in the distributions of the stellar and gas
components. The lack in the observed velocity element, that implies
absence of rotation, and contradicts with the consequences of
exponential profile of the stellar component, may also be a result
of the gravitational interactions.} {} 

\maketitle

\section{Introduction}

SMC is classified as an irregular dwarf galaxy, characterized by a
pronounced central feature of a Bar and an eastern extension called
the Wing. There has been a lot of investigation carried out studying
the distribution of different stellar populations of this galaxy.
Initially, \cite{freeman} has calculated the scale length of the
exponential component of the SMC from B-magnitude photometry. He has
found a$^{-1}$=0.63~kpc. A long time later, \cite{gardiner} have
found for a northern direction an exponential profile with
a$^{-1}$=1.2~kpc, describing well the distribution up to r=6kpc.
However this value was calculated over the observed projected radial
distribution assuming a spherically symmetric halo structure.

\cite{morgan} have carried out a spectroscopic survey of carbon
stars in the outer parts of the SMC. Regarding the spatial
distribution of these stars, they conclude that at most surface
densities the SMC appears elliptical with major axis parallel to the
Bar, but there is a significant northward distortion of the outer
contours. The diameter of the outermost contour is
$\sim$10$^{o}$-12$^{o}$. The authors also underline the existence of
a spiral-arm-like extension of the carbon stars distribution
southwards at (projected) distances larger than 4$^{o}$ ($\sim$4
kpc) from the optical centre, although it's origin is not clear.

Counts of sources towards the Magellanic Clouds from DENIS
near-infrared survey have been studied by \cite{cioni}. Their
investigation was based on the differentiation in the (I-J, I)
Colour Magnitude Diagram (CMD) in order to distinguish between three
groups of objects with different mean ages. The spatial distribution
of the three age groups is found quite different the youngest stars
exhibit an irregular structure while the older stars are smoothly
and regularly distributed. The AGB and RGB stars are characterised
by a regular, but double peaked, structure, an offset from the HI
distribution, and a mean age difference of the two maxima.

\cite{maragoudaki} have studied the spatial distribution of the SMC
stellar population according to their age, based on optical data.
They have shown that the older stellar population shows a rather
regular and smooth distribution, while the youngest stellar
component (age less than 8$\times$10$^{6}$~yr) appears mainly along
the northeast-southwest direction forming the Bar.

\cite{stanimirovic} in a detailed study of the HI kinematics
conclude that the HI velocity field shows a large velocity gradient
from the south west to the north east. The isovelocity contours of
this velocity field show some symmetry, suggestive of a differential
rotation. Some large-scale distortions in this velocity field are
easily visible but could be related to positions of several
supergiant shells. In the same article the authors conclude that, in
contrast to HI distribution the old stellar populations appear to
have a spheroidal spatial distribution and a total absence of
rotation.

In order to further investigate this discrepancy between the gas and
stellar component distribution, we have studied the distribution of
the various stellar components of the SMC from near IR surveys. We
know that IR observations characterize the mass of the galaxy better
than those in shorter wavelengths where absorption masks a large
part of the stellar component. A detailed study of the stellar
component as it is revealed from 2MASS is given in section 2. A
comparison with available radio data is given in section 3,
implementing the same approach used for the near IR data. Discussion
and conclusions are given in section 4.

\section{2MASS data}

The near-infrared (NIR) images were obtained by the Two-Micron
All-Sky Survey (2MASS), an ongoing effort to map the entire sky at
J-band (1.25  microns), H-band (1.65 microns) and K-band (2.17
microns) wavelengths. The 2MASS survey is led by the University of
Massachusetts, with all data and images processed at Caltech's
Infrared Processing and Analysis Centre (IPAC). The survey utilizes
two nearly identical 1.3-meter telescopes located at Mount Hopkins
(Arizona) and at Cerro Tololo (Chile). While the pixel size is 2.0
arcseconds, the survey strategy of over-sampling yields an effective
resolution of about 1 arcsecond. Exposure times are 7.8 seconds.
2MASS has uniformly scanned the entire sky in three near-infrared
bands to detect and characterize point sources brighter than about 1
mJy in each band, with signal-to-noise ratio (SNR) greater than 10.
This has achieved an 80,000-fold improvement in sensitivity relative
to earlier surveys. The data used here are from the 2MASS All-Sky
Point Source Catalog (PSC) (fp\_psc), at IPAC Infrared Science
Archive (IRSA), Caltech/JPL
(http://irsa.ipac.caltech.edu/applications/Gator).

\begin{figure}
\centering
\includegraphics[angle=-90,width=8cm]{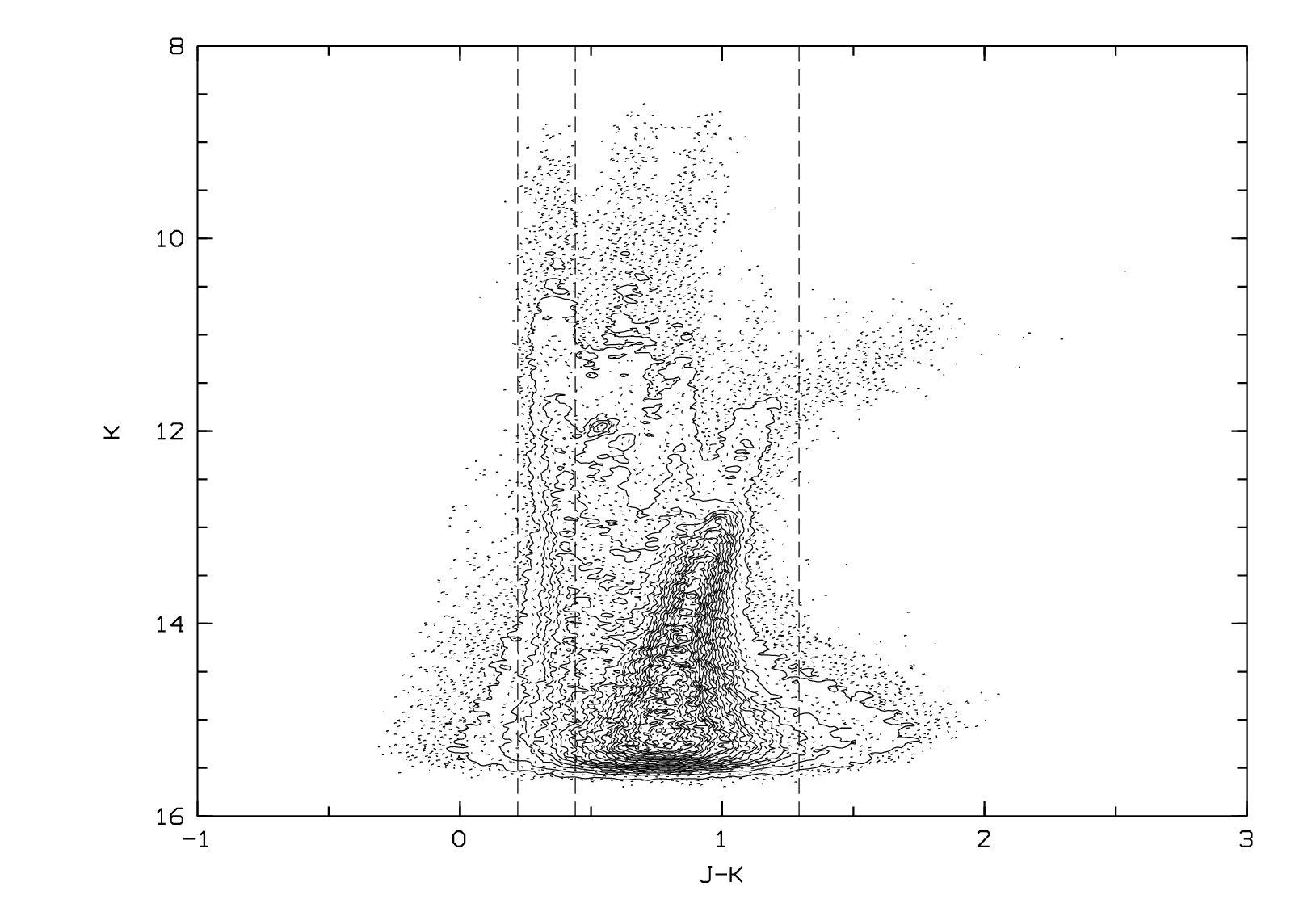}
\caption{Colour Magnitude Diagram K vs. J-K, of the SMC, from 2MASS}
\label{f01}
\end{figure}

IRSA gives choices for the origin of the default magnitudes and
uncertainties in each band. We have requested only sources with
satisfying quality of photometry for all J, H and K band. In order
to accomplish that we choose the case where: the default magnitude
is derived from a profile-fitting measurement made on the 1.3 sec
"Read\_2" exposures. The profile-fit magnitudes are normalized to
curve-of-growth-corrected aperture magnitudes. This is the most
common type in the PSC, and is used for sources that have no
saturated pixels in any of the 1.3 sec exposures.

The examined area of SMC is 6.9$^{o}\times$12$^{o}$. Data from a
small near by field of 2.1$^{o}\times$ 1$^{o}$ are also obtained in
order to be used later on for calculating the Galactic contribution.
The coordinates of this near by field are 47.5$\leq$R.A.$\leq$60 and
-68$\leq$Dec.$\leq$-67. It was selected to be close, but out of the
SMC field, in order to represent well the distribution of background
stars near SMC. The dimensions ware chosen to be big enough to make
statistics for the SMC field. The initial catalogues contain 359234
stars for the SMC area and 10444 stars for the small nearby field.
Total photometric uncertainty for 2MASS data is given by IRSA
(details at
http://www.ipac.caltech.edu/2mass/releases/allsky/doc/sec4\_5e.html).
Sources with total photometric uncertainty greater or equal to 0.2
mag for any of the J, H and K bands were excluded, providing more
reliable photometric measurements. Finally the catalogues used
contain 293330 stars for the SMC field and 8428 stars for the small
nearby field.

\subsection{Identifying Stellar populations}

Major stellar populations are identified based on matching features
of the observed colour-magnitude diagram with expected positions of
known populations. Based on Hiparchos data of SMC stars,
\cite{gavras} has determined the criteria for the different star
types on the K versus J-K CMD of the SMC. The A, B type stars are
found at -1 $<$ J-K $\leq$ 0.2 and K $\leq$ 15. The F and G type
have 0.2 $<$ J-K $\leq$ 0.45 and K $\leq$ 15. The K, M and faint
Carbon Stars meet the criteria of 0.45$\leq$ J-K $\leq$ 1.3 and K
$\leq$ 15 and finally Carbon stars have J-K $\geq$ 1.3 and K $\leq$
15. Applying the criteria above, to the 293330 stars for the SMC
field we find 2618 A, B type stars, 30042 of F and G type, 177471 K,
M and faint Carbon Stars and 4705 Carbon stars.

Two CMDs: the K vs J-K for the SMC area and the nearby field were
produced. Both diagrams are divided in a grid with the same number
of cells (22x16). The cells have dimensions that provide a fine
grid, while they include enough number of stars to make statistics.
The grids had the same number of cells and same starting and ending
points for each axis, in order to compare numbers of stars with same
K and J-K values. The cells of the nearby field were normalized to
the equal area of the cells of the SMC area and their stars
considered as background stars were subtracted randomly. The CMD of
K versus J-K for the SMC field after subtraction of the background
contribution is shown in Fig.\ref{f01}. The criteria set by
\cite{gavras} for J-K values are indicated by the dashed lines.

\begin{figure*}
\centering
\includegraphics[angle=0,width=6.25cm]{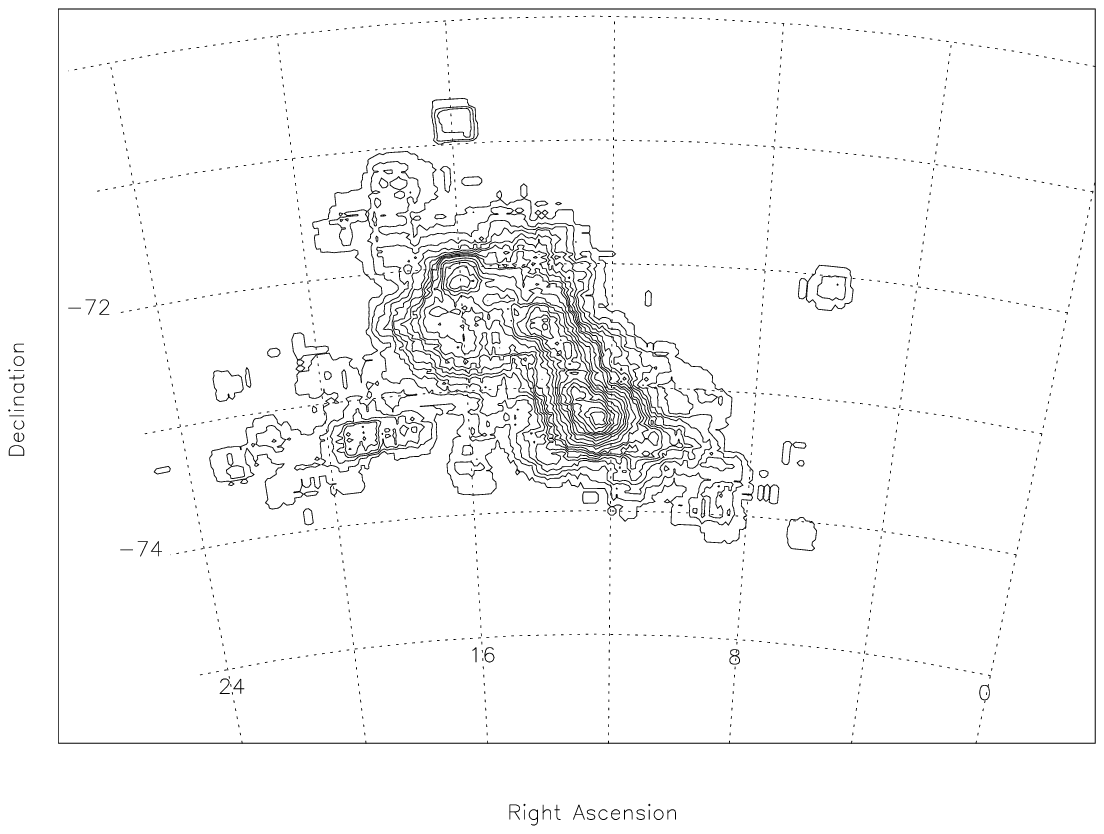}
\hspace{0.5cm}
\includegraphics[angle=0,width=6.25cm]{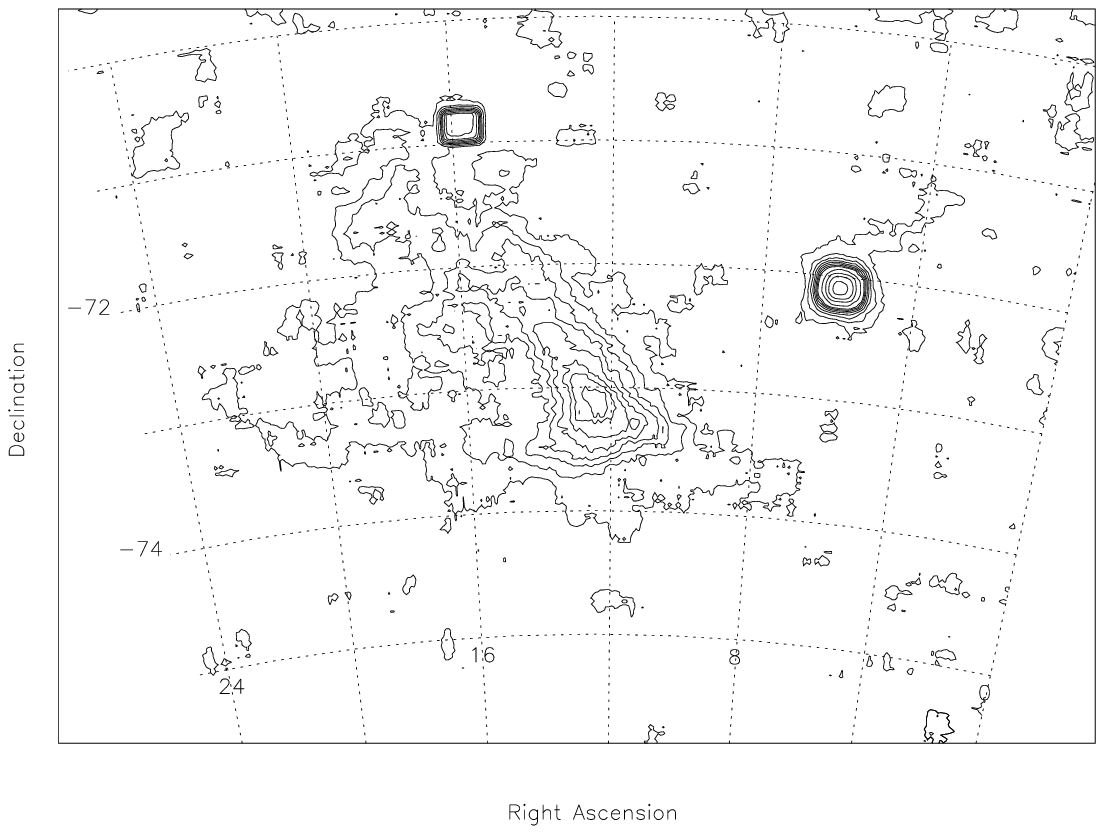}
\includegraphics[angle=0,width=6.25cm]{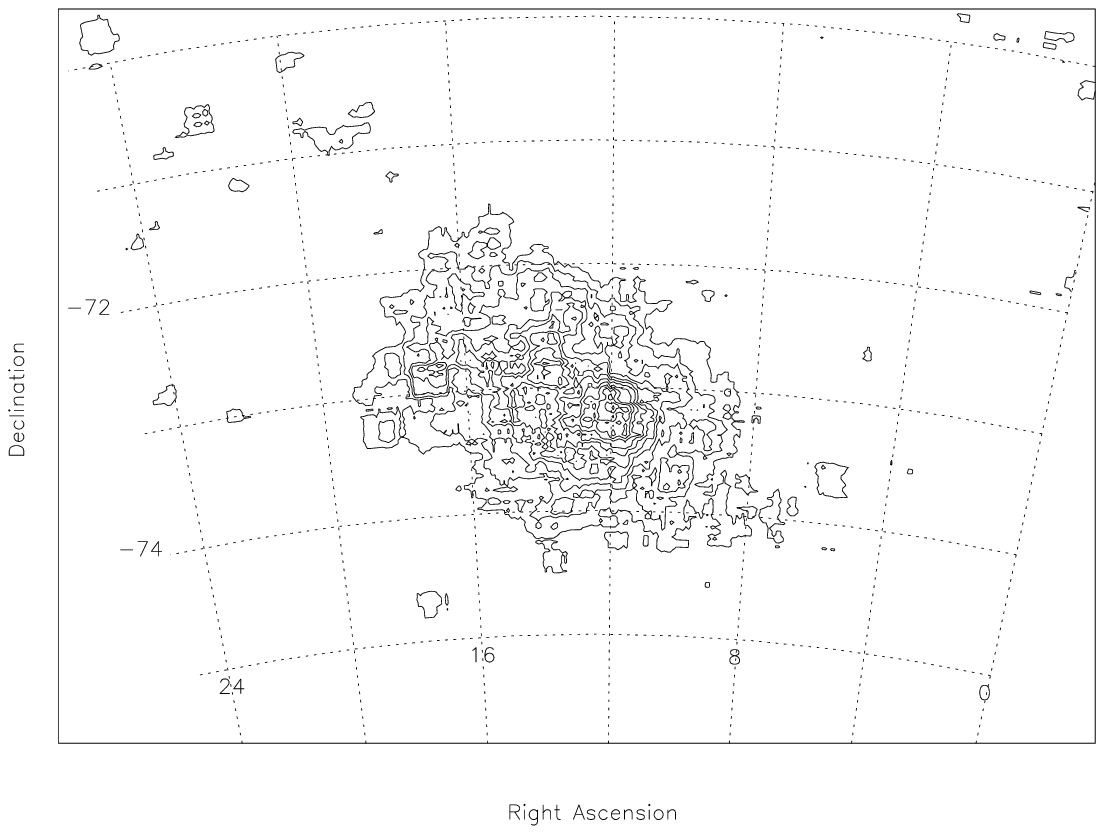}
\hspace{0.5cm}
\includegraphics[angle=0,width=6.25cm]{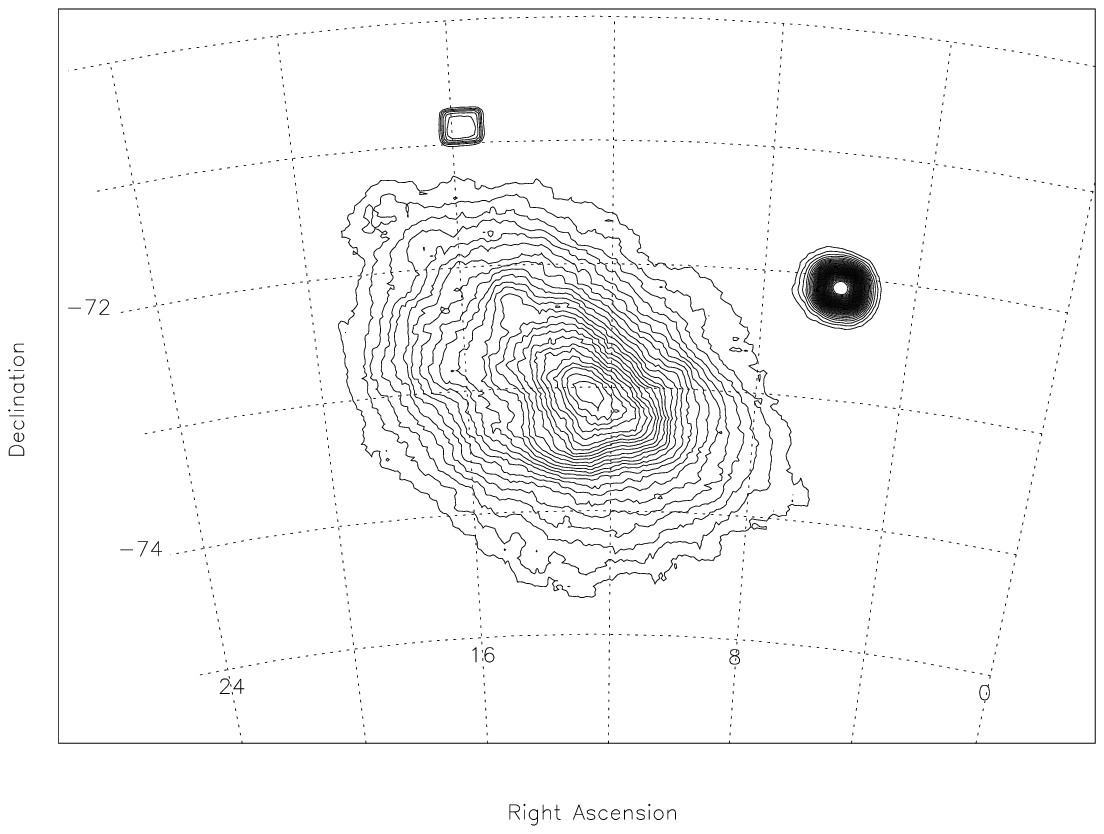}
\caption{Isopleths contour maps of the SMC different stellar
populations, from 2MASS. Top left: A and B type stars, top right: F
and G, bottom left: Carbon stars and bottom right: K, M and faint
Carbon.} \label{f02}
\end{figure*}

Since the different populations were defined, star counts were
performed in a grid of 288x500 pixels for the SMC region. The
isopleths contour maps of the SMC field that were produced for each
of the four cases can be seen in Fig.\ref{f02}. It is very
interesting to point out how well the old population (bottom right
in Fig.\ref{f02}) reveals the center of mass of this stellar
population.

\subsection{Distribution of late type stars}

\begin{figure*}
\centering
\includegraphics[angle=0,width=12cm]{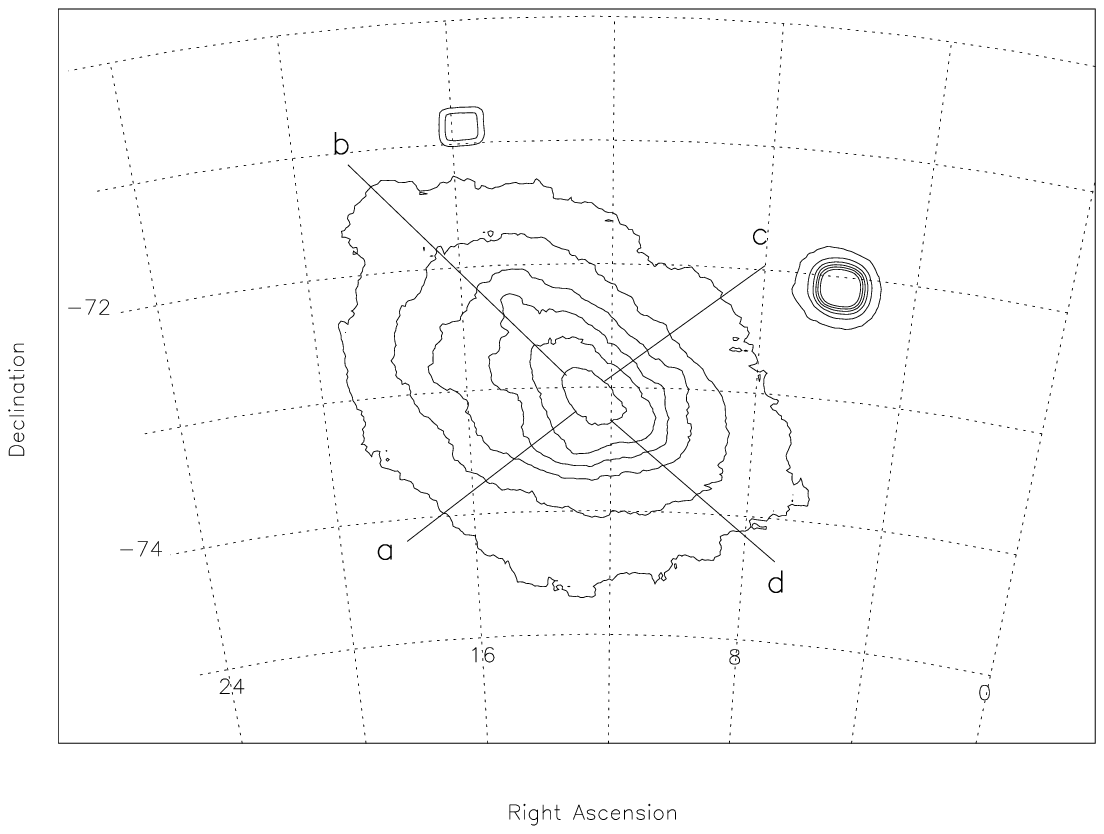}
\caption{Isopleths contour map of the SMC K, M and faint Carbon
Stars, from 2MASS. Lines indicate the selected directions. The
"major" axis of the galaxy is the one following the direction of the
Bar, from north-east to south-west, while the "minor" axis is
directing from south-east to north-west.} \label{f03}
\end{figure*}

In the case of K, M and faint Carbon stars the distribution reveals
the elliptical structure of a disk. The number density of the stars
versus the distance from the centre has been estimated. The
isopleths contour map of the SMC field was reproduced
(Fig.\ref{f03}) with bigger steps between the contours in order to
allow more accurate selection of the image points that will be taken
under consideration. The outermost contour is the same as in
Fig.\ref{f02} and will be the one specifying the diameter of each
structure. We selected 4 directions a, b, c and d along the "minor"
and "major" axis to derive their density profiles. The "major" axis
is the one following the direction of the Bar, from north-east to
south-west, while the "minor" axis is directing from south-east to
north-west.

\begin{figure*}
\centerline{\hbox{
\includegraphics[angle=0,width=6cm]{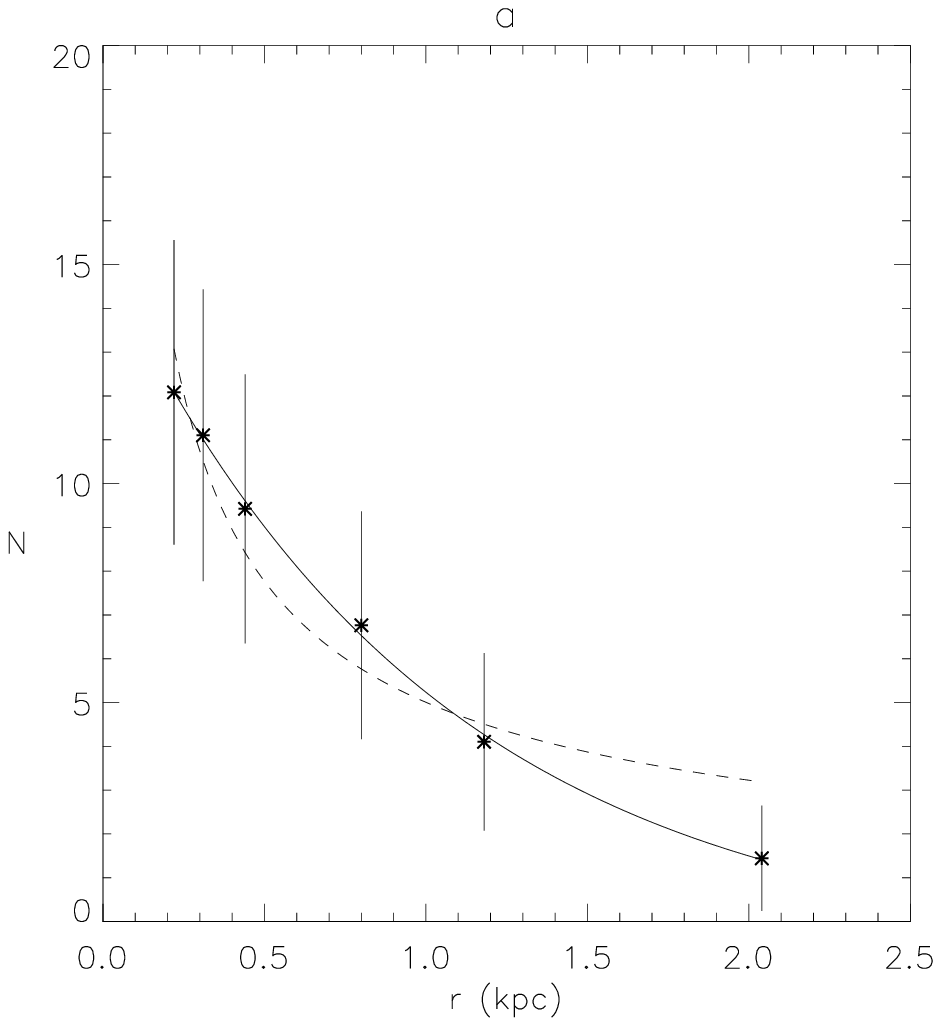}
\includegraphics[angle=0,width=6cm]{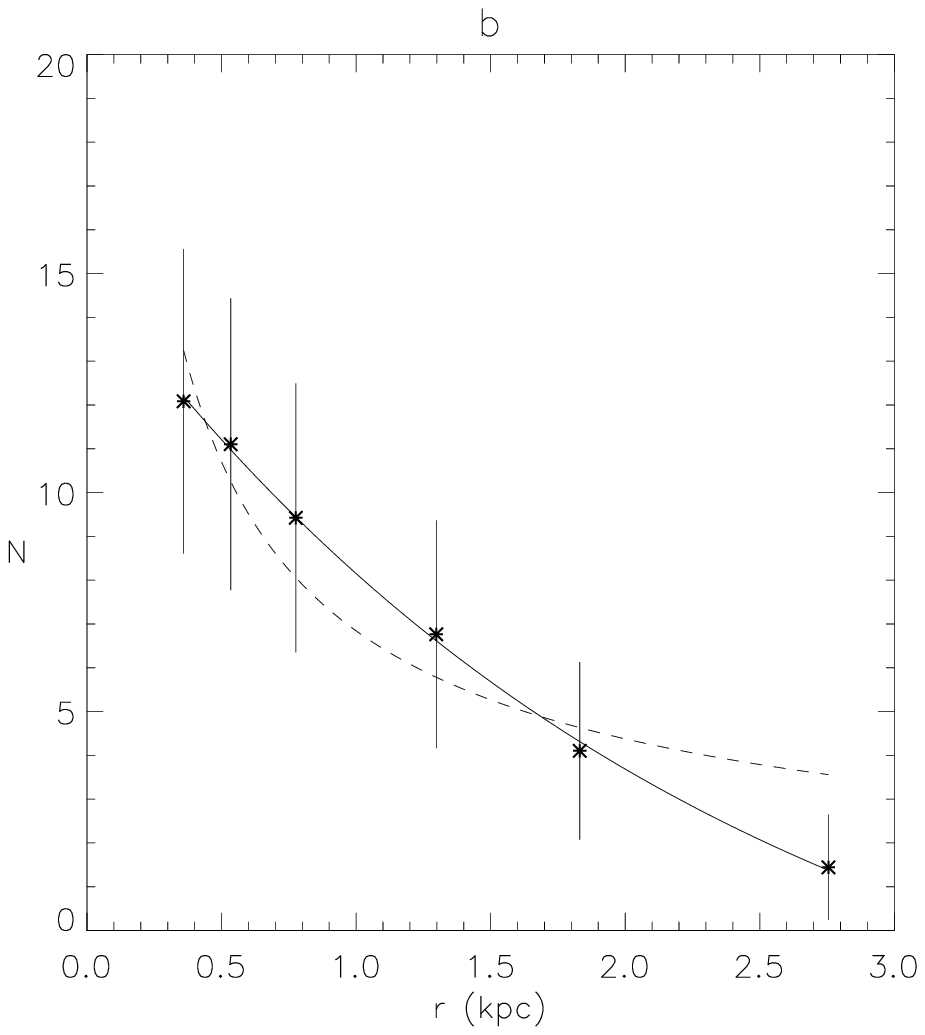}
}}
\vspace*{0.25cm}
\centerline{\hbox{
\includegraphics[angle=0,width=6cm]{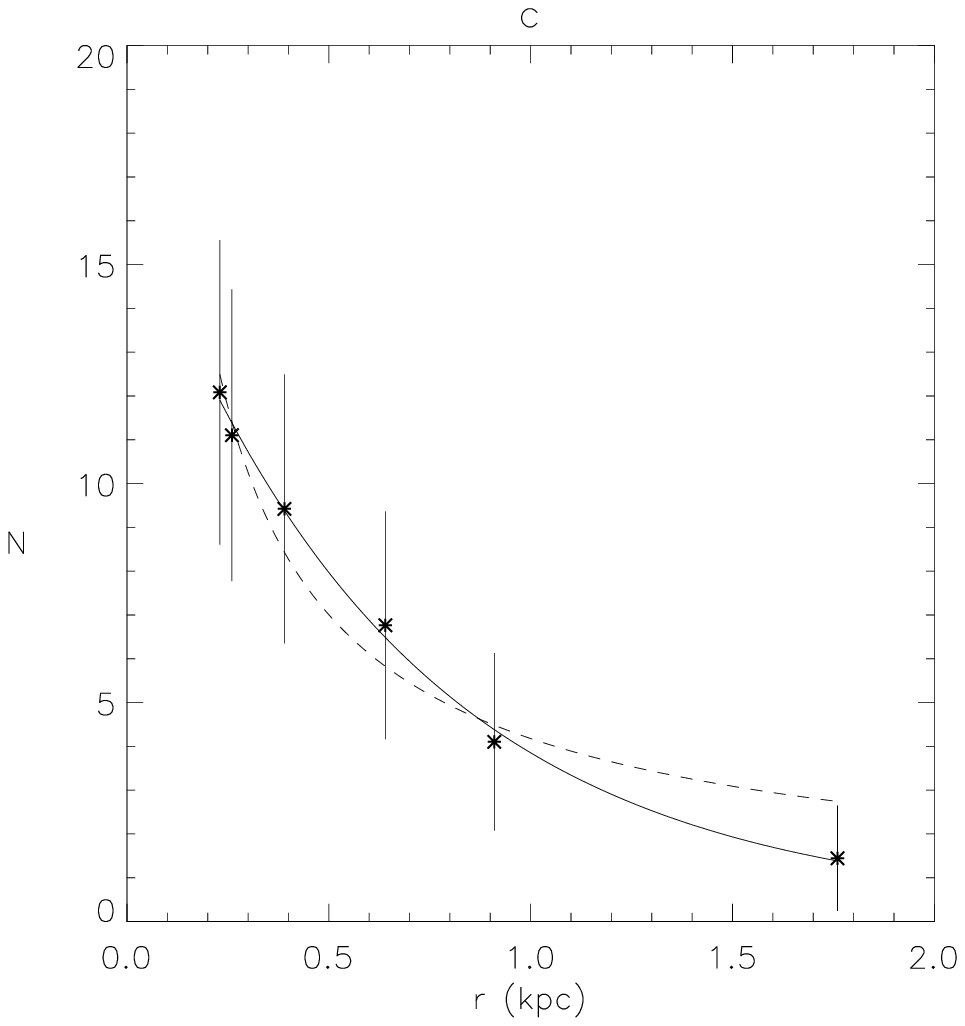}
\includegraphics[angle=0,width=6cm]{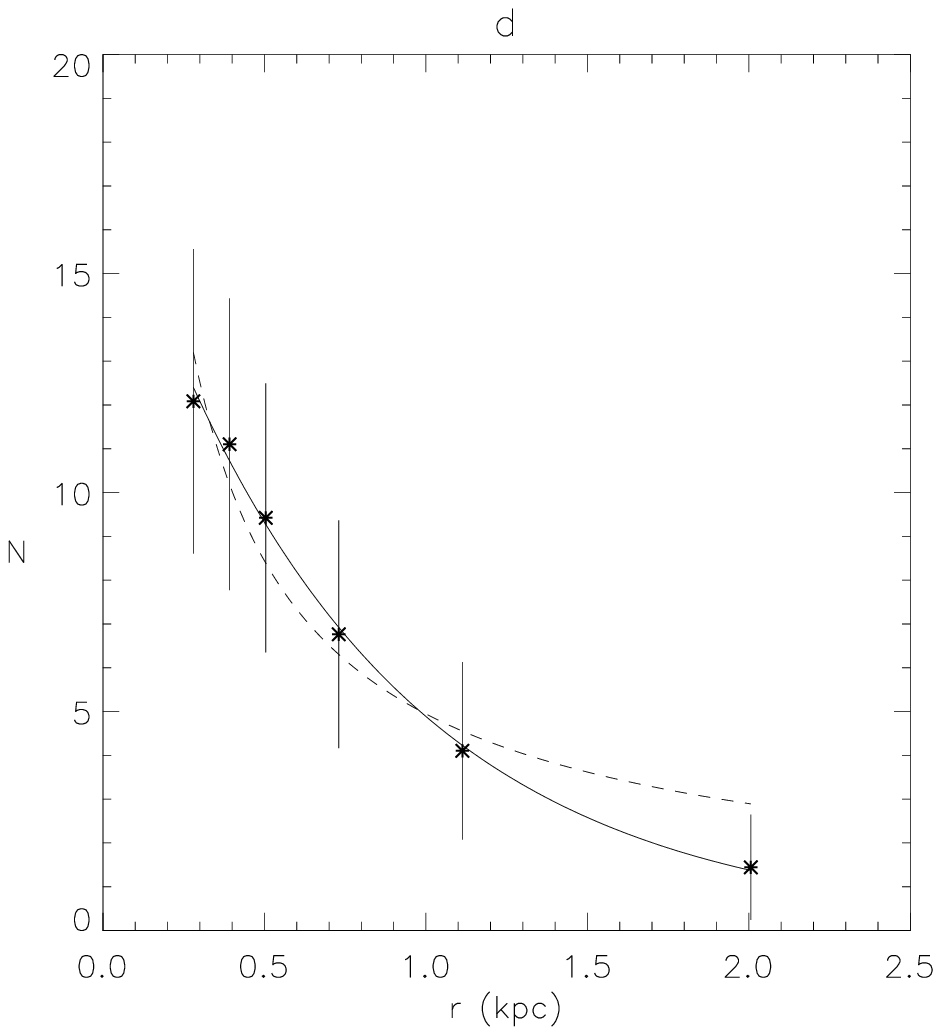}
}} \caption{Number of stars per pixel (N) vs. distance from the
centre (r) (star symbols), fitted with exponential curves (solid
lines) and power-law curves (dashed lines), for 2MASS data for the
a, b, c and d directions respectively.} \label{f04}
\end{figure*}

Fitting exponential curves to the diagrams of number density of
sources (N) versus (projected) distance from the centre of the
galaxy (r), with exponential functions was very satisfying. The best
fit functions are Y=15.9$\pm$2.2*exp(-1.0$\pm$0.5*X)-0.8$\pm$3.1,
Y=19.7$\pm$6.8*exp(-0.4$\pm$0.3*X)-4.7$\pm$8.1,
Y=16.6$\pm$1.8*exp(-1.5$\pm$0.6*X)+0.3$\pm$2.0,
Y=17.8$\pm$2.0*exp(-1.3$\pm$0.5*X)-0.1$\pm$2.1, for the directions
a, b, c and d respectively. The values of a$^{-1}$ are 1.0, 2.5, 0.7
and 0.8 derived for each direction respectively, corresponding to a
mean value of a$^{-1}$= 1.25~kpc. For the power-law scenario, the
best fit functions are Y=5.0$\pm$0.5*X$^{(-0.6\pm0.1)}$,
Y=6.8$\pm$0.5*X$^{(-0.6\pm0.1)}$, Y=4.2$\pm$0.6*X$^{(-0.8\pm0.1)}$,
Y=5.0$\pm$0.5*X$^{(-0.8\pm0.1)}$, for the directions a, b, c and d
respectively. The chi-square parameters are 0.04, 0.03, 0.09 and
0.11 in the case of exponential fitting curves for the directions a,
b, c and d respectively. The corresponding values for the power-law
case are 1.6, 2.4, 1.0 and 2.4, indicating that exponential fitting
represents better the data. Consequently the mass distribution
derived here is described by an exponential disk as illustrated in
Fig. \ref{f04} with the best fitting curves. Summing up, 2MASS
reveals an exponential profile for the older stellar population with
a (projected) diameter $\sim$ 3.8 kpc for the "minor" axis and
$\sim$ 4.8 kpc for the "major" axis.  The center of the elliptical
distribution is best determined by the right bottom map of Fig.
\ref{f02}, at RA: 0$^{h}$ 51$^{m}$, Dec: -73$^{o}$ 7$\arcmin$.2,
that represents the K, M and faint carbon stars.

The exponential disk model for K, M and faint Carbon stars, has to
face the fact that the values of a$^{-1}$ are not found the same for
the semi-axis's, and the fact that exponential along direction (d)
is measured to be "steeper" than the exponential along direction
(a). The observed asymmetry might be explained by the inclination of
the SMC,  although deprojection procedure is not clear enough in the
case of this galaxy. However further investigation should put light
on this issue.

\section{Radio data}

\begin{figure*}
\centering
\includegraphics[angle=0,width=12cm]{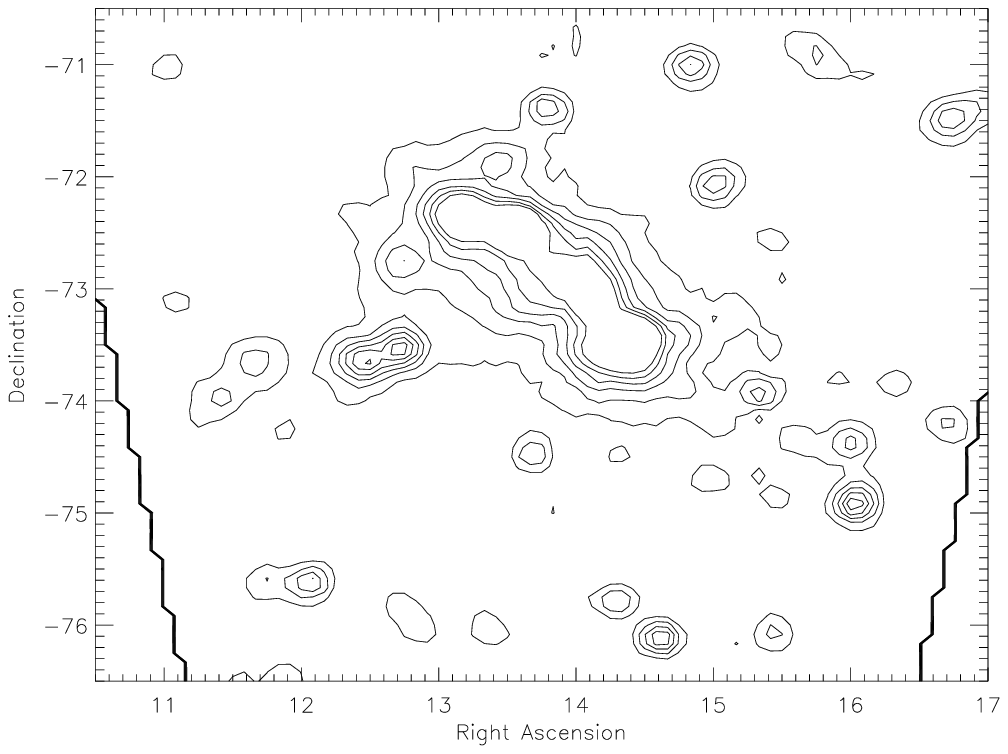}
\caption{Isodensity contour map of the SMC radio data.} \label{f05}
\end{figure*}

\begin{figure*}
\centerline{\hbox{
\includegraphics[angle=0,width=6cm]{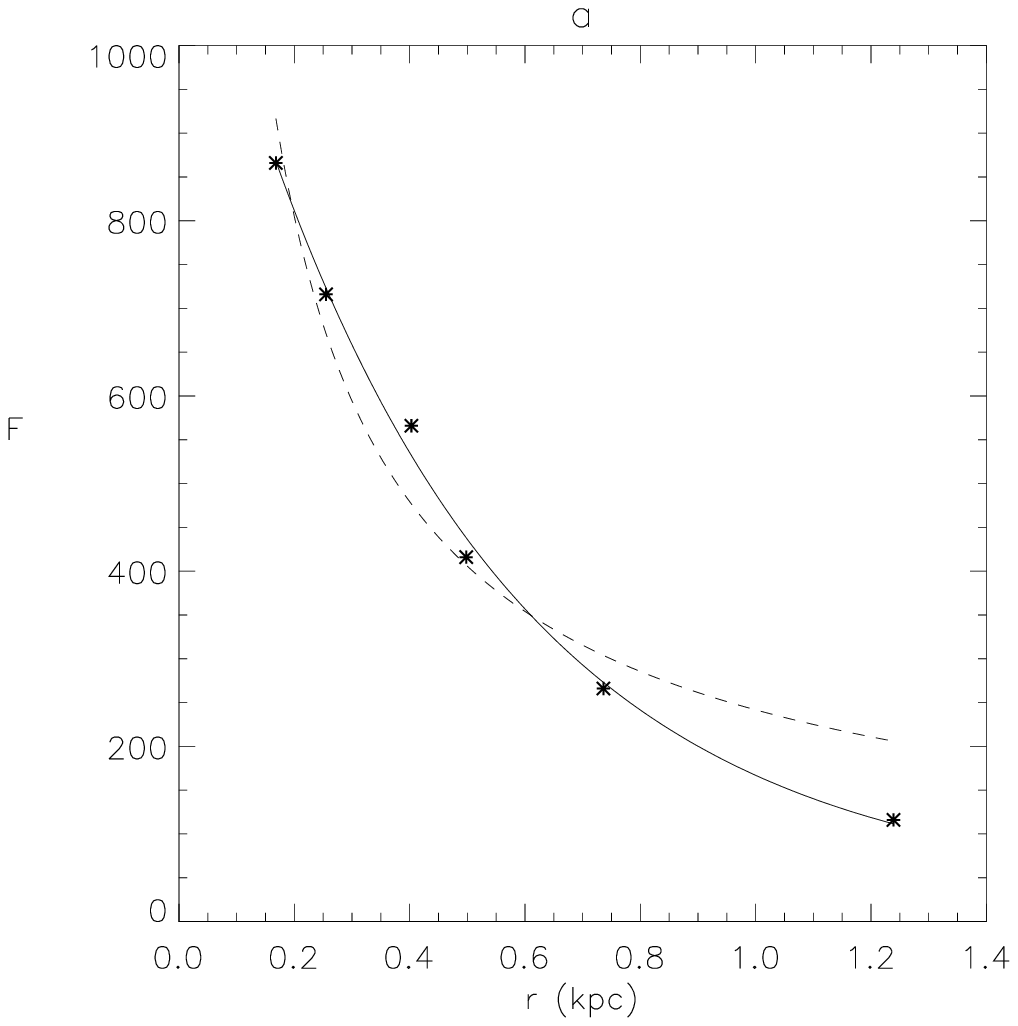}
\includegraphics[angle=0,width=6cm]{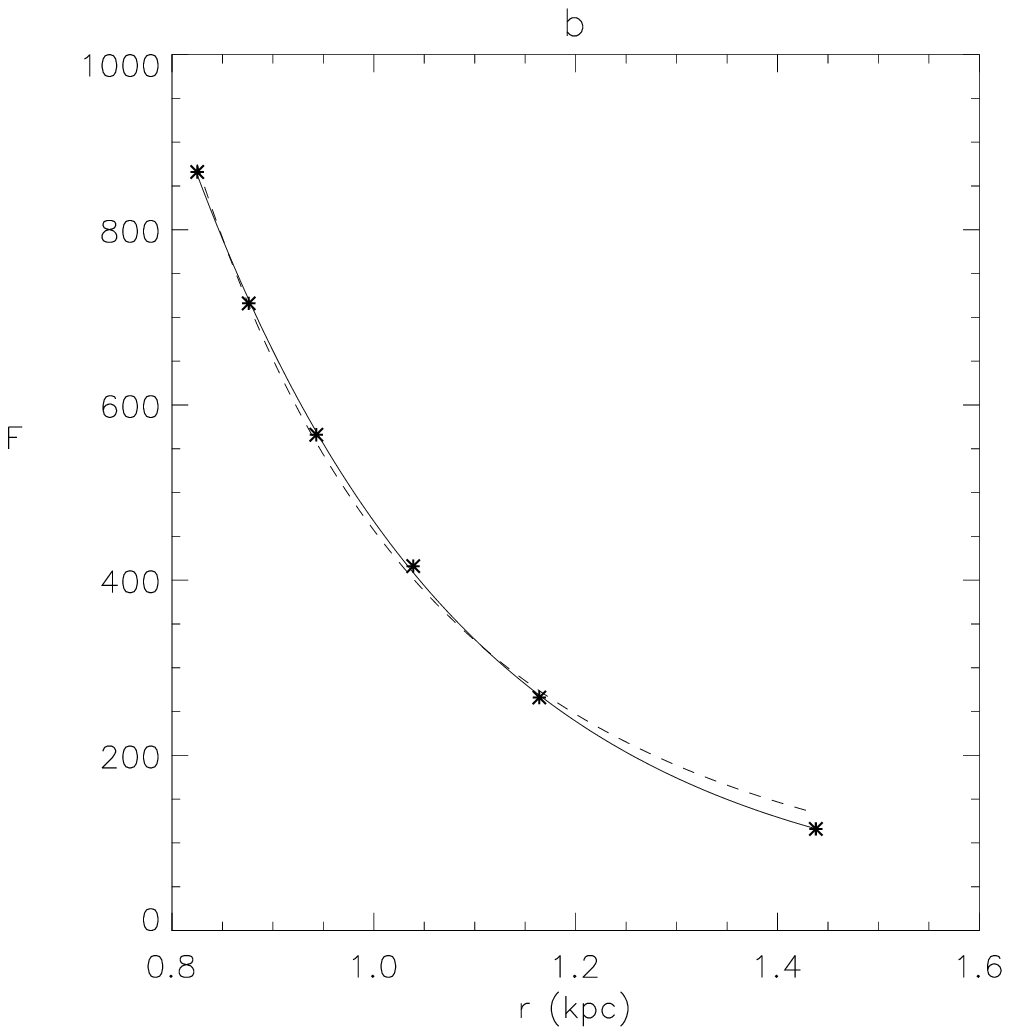}
}}
\centerline{\hbox{
\includegraphics[angle=0,width=6cm]{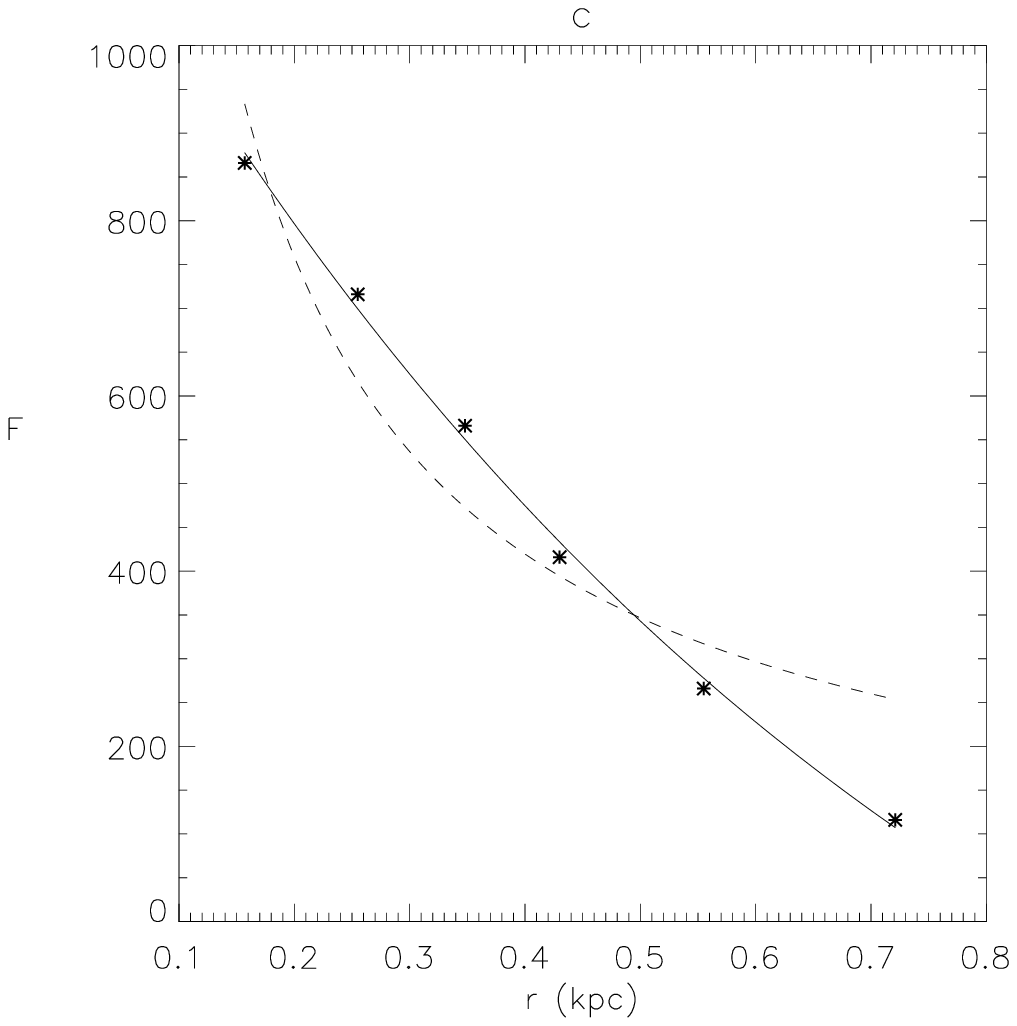}
\includegraphics[angle=0,width=6cm]{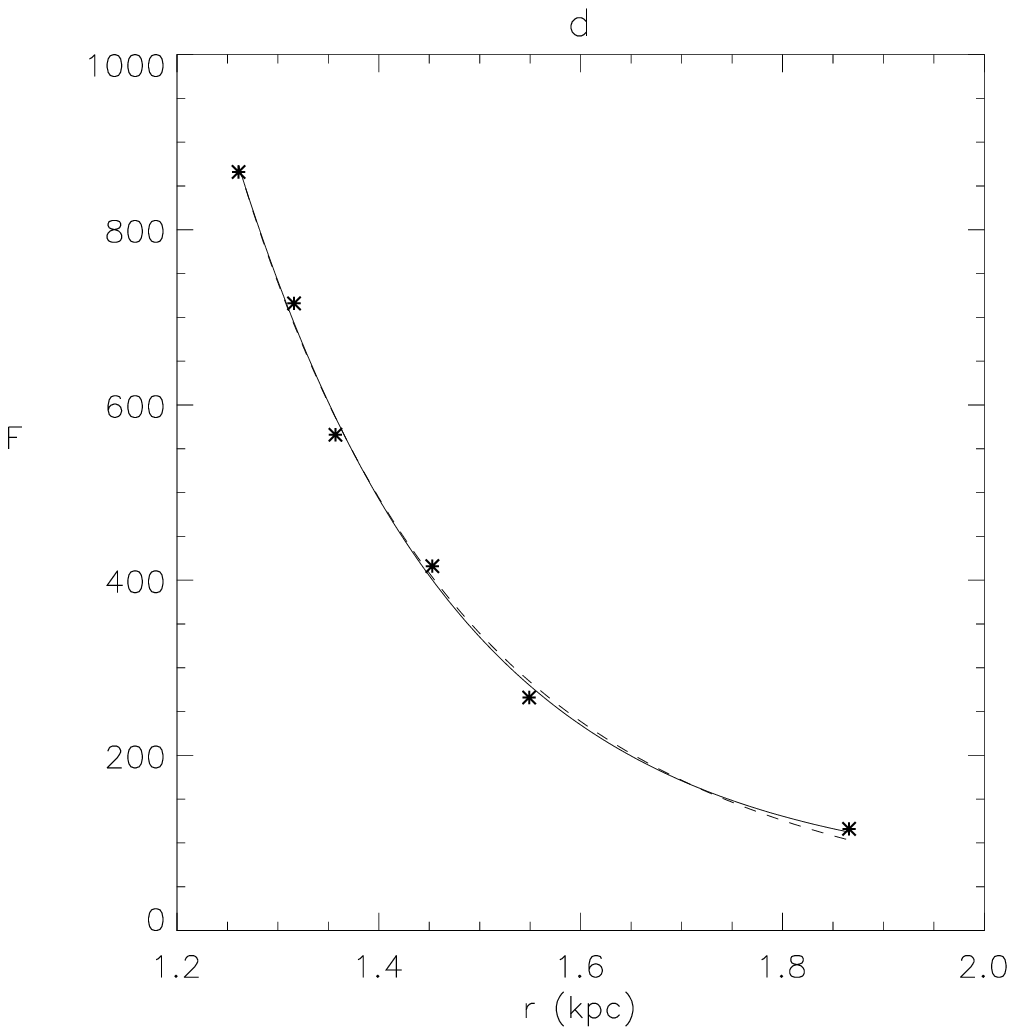}
}} \caption{Flux density (F) versus distance from the centre (r)
(star symbols), fitted with exponential curves (solid lines) and
power-law curves (dashed lines), for radio data for the a, b, c and
d directions respectively.} \label{f06}
\end{figure*}

A radio image of SMC at 1400 GHz (21cm) was obtained from Uli Klein
(private communication) based on the data that were derived from the
work by \cite{haynes}. The data represent well the extended radio
envelope of the SMC, well beyond any site of obvious active star
formation and are appropriate for investigating the gas distribution
at the SMC. The radial profile of this image is presented in Fig.
\ref{f05}. Using the previous methodology we created diagrams of
flux density (F) versus distance from the centre (r), and fitted
data with exponential and power-law curves in order to reveal the
most efficient mathematical description of this distribution. The
best fit exponential functions we found are
Y=1208.1$\pm$2.0*exp(-2.2$\pm$0.01*X)+30.7$\pm$1.9,
Y=17243.6$\pm$270.6*exp(-3.7$\pm$0.02*X)+28.2$\pm$2.1,
Y=1801.9$\pm$21.2*exp(-1.33$\pm$0.02*X)-585.6$\pm$14.6, and
Y=248641.0$\pm$6690.8*exp(-4.5$\pm$0.02*X)+60.13$\pm$1.6, for the
directions a, b, c and d respectively. The values of a$^{-1}$ are
0.45, 0.27, 0.75 and 0.22 derived for each direction respectively,
corresponding to a mean value of a$^{-1}$= 0.42~kpc. For the
power-law case, the best fit functions are
Y=241.6$\pm$0.5*X$^{(-0.8\pm0.001)}$,
Y=456.9$\pm$0.5*X$^{(-3.4\pm0.01)}$,
Y=191.7$\pm$0.5*X$^{(-0.9\pm0.002)}$,
Y=3087.4$\pm$10.8*X$^{(-5.5\pm0.01)}$, for the directions a, b, c
and d respectively. The chi-square parameters are 590.0, 390.8, 33.8
and 471.9 in the case of exponential fitting curves for the
directions a, b, c and d respectively, whereas the corresponding
values for the power-law case are 5542.13, 11285.9, 188.0 and 419.5.
The comparison of the two sets of parameters allow us to suggest
that exponential fitting represents better the distribution. We
notice however, that in for direction (d) power-law is more
efficient the Consequently the mass distribution derived here is
described by an exponential disk as illustrated in Fig. \ref{f06}
with the best fitting curves.

These led us to a mean value of a$^{-1}$= 0.42~kpc for the radio
data. The centre is found at RA: 0$^{h}$ 55$^{m}$ 7$^{s}$.2, Dec:
-72$^{o}$ 47$\arcmin$ 59$\arcsec$. The HI exponential  profile has a
scale length of 0.45 kpc along the south-east direction (a), 0.27
kpc along the north-east direction (b), 0.75 kpc along the to
north-west direction (c) and 0.22 kpc along the south-west direction
(d). These coordinates of the center are in accordance with the
results of \cite{stanimirovic} derived from higher resolution data,
and confirms that the differences found here between the central
coordinates of the stellar and the gas component are real.

\section{Discussion \& Conclusions}

The distribution of the stellar components in the near IR (2MASS)
has been used to investigate the radial distribution of the stellar
mass in the SMC. The isopleths in near IR wavelengths are expected
to represent more accurately the distribution of stars since the
observations are less affected by absorption.

In the case of the 2MASS data we set the criteria for collecting the
different stellar populations and present their spatial
distributions. It has been found that exponential profile fits
better the distribution of the older population of the SMC. In Fig.
\ref{f02} the bottom right map of K, M and faint carbon stars,
reveals extremely accurately the center of the mass distribution of
the oldest stellar population of the SMC, the coordinates of this
center are found to be the same for all cases, independently the age
of population they represent.

This centre of the isopleths is found to be RA: 0$^{h}$ 51$^{min}$,
Dec: -73$^{o}$ 7$\arcmin$ and the mean value for scale length is
a$^{-1}$=1.25~kpc.

\subsection{The center offset}

The important characteristic of the distribution of the older
stellar population of the SMC is found to be the offset of the
center of their distribution from the centre of the distribution
found by HI data  \citep{stanimirovic}. In a more recent work,
\cite{piatek} have calculated the kinematical centre of the SMC at
RA: 0$^{h}$ 52$^{m}$.8, Dec: -72$^{o}$ 30$\arcmin$ from HST data
being in good agreement with our calculations. They also explain
that the presence of a Bar or a strong tidal disturbance can infer
extra streaming motions to a system and thus old and young
populations can have distinct kinematics. The strong interactions of
the SMC with the LMC and the MilkyWay that have probably contributed
to the formation of the Bar \citep{maragoudaki} could also have as a
result the observed difference in the centres of the distributions
of the stellar and the gas component as well as the small radial
velocity gradient among SMC stars \citep{bekki}, that contradicts
with the exponential profile of the stellar component.

\cite{stanimirovic} suggest that according to SMC´s rotation curve a
dark matter halo is not needed to explain it´s dynamics. However
reconsidering the values of the center of the SMC, adopting the
center of the mass derived above, when calculating rotational
velocity profiles may be important considering the issue of dark
matter presence in this galaxy. SMC is a dwarf irregular galaxy,
which are expected to be dark matter dominated (\citealt{cote};
\citealt{begum}; \citealt{strigari}). However, rotational velocities
do not support this case for SMC, probably because they have been
affected by the interactions of the Magellanic system with the Milky
way.

\subsection{The scale length values}
As mentioned before, \cite{freeman} has investigated the structure
of the SMC through bright stars distribution and has found that
their elliptical component is characterised by a scale length of
a$^{-1}$=0.63 kpc. We find an exponential profile for the older
stellar population with  a$^{-1}$=1.25~kpc, while the gas
exponential disk has a$^{-1}$= 0.42~kpc. These three values
represent well the structure of three different galactic components
and are in accordance with the fact that gas (radio data) is more
concentrated on a thinner disk, younger stars (best represented by
B-magnitude distribution) lay on a thicker disk and finally oldest
stars reach bigger distances from the center.

\section{Acknowledgements}

The authors would like to thank the University of Athens (ELKE) for
partial financial support. The project is co-funded by the European
Social Fund and National Resources-(EPEAEK II) PYTHAGORAS II.





\begin{thebibliography}{}

\bibitem[\protect\citeauthoryear{Begum et al.}{2006}]{begum}
Begum, A., Chengalur, J. N., Karachentsev, I. D., Kaisin, S. S., \&
Sharina, M. E. 2006, MNRS, 365, 1220

\bibitem[\protect\citeauthoryear{Bekki \& Chiba}{2008}]{bekki}
Bekki, K., \& Chiba, M. 2008, astro-ph, 0806.4657

\bibitem[\protect\citeauthoryear{Cioni et al.}{2000}]{cioni}
Cioni, M.-R. L., Habing, H. J., \& Israel, F. P. 2000, A\&A, 358, L9

\bibitem[\protect\citeauthoryear{Cote et al.}{2000}]{cote}
Cote, S., Carignan, C., \& Freeman, K. C. 2000, AJ, 120, 3027

\bibitem[\protect\citeauthoryear{Freeman}{1970}]{freeman}
Freeman, K. C. 1970, ApJ, 160, 811

\bibitem[\protect\citeauthoryear{Gardiner \& Hawkins}{1991}]{gardiner}
Gardiner, L. T., \& Hawkins, M. R. S. 1991, MNRAS, 251, 174

\bibitem[\protect\citeauthoryear{Gavras}{2003}]{gavras}
Gavras, P. 2003, MSc thesis University of Athens

\bibitem[\protect\citeauthoryear{Haynes et al.}{1986}]{haynes}
Haynes, R. F., Klein, U., Wielebinski, R., \& Murray, J. D. 1986,
A\&A, 159, 22

\bibitem[\protect\citeauthoryear{Maragoudaki et al.}{2001}]{maragoudaki}
Maragoudaki, F., Kontizas, M., Morgan, D. H., et al.
2001, A\&A, 379, 864

\bibitem[\protect\citeauthoryear{Morgan \& Hatzidimitriou}{1995}]{morgan}
Morgan, D. H., \& Hatzidimitriou, D. 1995, A\&AS, 113, 539

\bibitem[\protect\citeauthoryear{Piatek et al.}{2007}]{piatek}
Piatek, S., Pryor, C., \& Olszewski, E. W. 2007, astro-ph, 0712.176

\bibitem[\protect\citeauthoryear{Stanimirovic et al.}{2004}]{stanimirovic}
Stanimirovic, S., Staveley-Smith, L., \& Jones, P. A. 2004, ApJ,
604, 176

\bibitem[\protect\citeauthoryear{Strigari et al.}{2008}]{strigari}
Strigari, L. E., Koushiappas, S. M., Bullock, J. S., et al.
2008, ApJ, 678, 614

\end{thebibliography}
\end{document}